\documentclass[aps,prl,twocolumn,superscriptaddress]{revtex4}%
\usepackage{amsfonts}
\usepackage{amsmath}
\usepackage{amssymb}
\usepackage{graphicx}
\usepackage{xcolor}
\usepackage{epstopdf}
\usepackage{bm}%
\begin{document}
\title{Giant Inverse Rashba-Edelstein Effect: Application to Monolayer OsBi$_{2}$}
\author{Rui Song}
\affiliation{HEDPS, Center for Applied Physics and Technology and School of Physics, Peking
University, Beijing 100871, China}
\affiliation{HEDPS, Center for Applied Physics and Technology and School of Engineering,
Peking University, Beijing 100871, China}
\affiliation{Anhui Key Laboratory of Condensed Matter Physics at Extreme Conditions, High
Magnetic Field Laboratory, HFIPS, Anhui, Chinese Academy of Sciences, Hefei,
230031, China}
\affiliation{Institute of Applied Physics and Computational Mathematics, Beijing 100088, China}
\author{Ning Hao}
\email{haon@hmfl.ac.cn}
\affiliation{Anhui Key Laboratory of Condensed Matter Physics at Extreme Conditions, High
Magnetic Field Laboratory, HFIPS, Anhui, Chinese Academy of Sciences, Hefei,
230031, China}
\author{Ping Zhang}
\email{zhang_ping@iapcm.ac.cn}
\affiliation{School of Physics and Physical Engineering, Qufu Normal University, Qufu
273165, China}
\affiliation{HEDPS, Center for Applied Physics and Technology and School of Engineering,
Peking University, Beijing 100871, China}
\affiliation{Institute of Applied Physics and Computational Mathematics, Beijing 100088, China}
\affiliation{Beijing Computational Science Research Center, Beijing 100084, China}

\begin{abstract}
We propose that the hybridization between two sets of Rashba bands can lead to
an unconventional topology where the two Fermi circles from different bands
own in-plane helical spin textures with the same chiralities, and possess
group velocities with the same directions. Under the weak spin injection, the
two Fermi circles both give the positive contributions to the spin-to-charge
conversion and thus induce the giant inverse Rashba-Edelstein Effect with
large conversion efficiency, which is very different from the conventional
Rashba-Edelstein Effect. More importantly, through the first-principles
calculations, we predict that monolayer OsBi$_{2}$ could be a good candidate
to realize the giant inverse Rashba-Edelstein Effect. Our studies not only
demonstrate a new mechanism to achieve highly efficient spin-to-charge
conversion in spintronics, but also provide a promising material to realize it.

\end{abstract}
\maketitle

The control and manipulation of the spin-charge interconversion plays a
critical role in modern spintronics \cite{PRM-spin-hall,spintronics-2004}. In
the two-dimensional system, the (inverse) Edelstein effect has gained much
attention since its potential application in spintronics
devices\cite{Edelstein-1989,PRL-2007-metal,NC-2013-metal,APL-2015-metal1,APL-2015-metal2,PRB-2016-metal3,PRB-2016-metal4,PRL-2015-metal,PRL-metal-2016,Edelstein-1,Edelstein-2,NM-1,PRL-Sn-2016}%
. In the direction of inverse Edelstein effect (IEE), a pure spin current
$j_{s}$ through the system generates a transverse charge current $j_{c}$, and
the Edelstein effect (EE) describes the inverse process. In both cases, the
conversion efficiency is defined by $\lambda_{(I)EE}=j_{c}/j_{s}$, which
largely measures the merits of a physical system.

The microscopic mechanism of both EE and IEE requires the presence of the
spin-orbit coupling (SOC), which results in the specific spin-momentum locked
electronic band structures. Thus, the usual candidate physical systems include
the metallic
heterostructure\cite{PRL-2007-metal,NC-2013-metal,APL-2015-metal1,APL-2015-metal2,PRB-2016-metal3,PRB-2016-metal4,PRL-2015-metal,PRL-metal-2016}
and the topological
insulators\cite{PRL-2007-TI,PRB-2007-TI,RMP-2010-TI,RMP-2011-TI,PRL-Sn-2016,PRB-TI-2017,IEE-TI-2018}%
. In the metallic heterostructure, the spacial inversion asymmetry lifts the
spin degeneracy and gives rise to the Rashba SOC\cite{Rashba}. However, the
opposite spin textures of the two lifted bands give the partial compensation
of the contributions to the spin-to-charge conversion and suppresses the
efficiency $\lambda_{IEE}$\cite{PRL-Sn-2016}, and searching for materials with
strong Rashba SOC coupling becomes an alternative way to increase
$\lambda_{IEE}$ in this
situation\cite{NM-2011-SOC,Giant-Rashba1,Giant-Rashba2,Giant-Rashba3,Giant-Rashba4,Micro-theory,PRB-theoretical-aspect}%
. Furthermore, the interfacial effects complicate the descriptions of the
electronic states beyond the standard Rashba model and limit the application
of the metallic heterostructure. In the topological insulator, the topological
surface state possesses the single Dirac cone structure, which can get rid of
of partial compensation effect in Rashba system\cite{PRL-Sn-2016}. However,
the concurrence of surface and bulk states and quantum confinement effect
always complicate TI-based systems beyond the
controllability\cite{Quan-confin-1}.

In this work, we propose the third kinds of system, which hosts a new
mechanism to achieve the giant IEE with large $\lambda_{IEE}$. The new system
has two spin-lifted bands with the identical spin textures, \textit{i.e.},
unconventional Rashba bands, which is very different from the conventional
Rashba bands with opposite spin textures. We first construct a simple and
generic model to describe such unconventional Rashba bands. We show that it
can induce strongly enhanced spin-to-charge conversion and possess the giant
IEE with large $\lambda_{IEE}$ according to the semi-classical Boltzmann
transport theory. The calculated spin-to-charge conversion efficiency
$\lambda_{IEE}$ is estimated to be ten times that of the conventional Rashba
system. More importantly, our first-principles calculations predict a simple
compound of monolayer OsBi$_{2}$ has such unconventional Rashba band
structure. The pure bulk states of a single material can overcome the
shortcomings of the complexity and sensitivity of the interfacial and surface
states in the metallic heterostructure and topological insulators,
respectively. These properties make monolayer OsBi$_{2}$ to be a promising
material to realize the giant IEE and to have potential application in spintronics.

\begin{figure}[pt]
\begin{center}
\includegraphics[width=1.0\columnwidth]{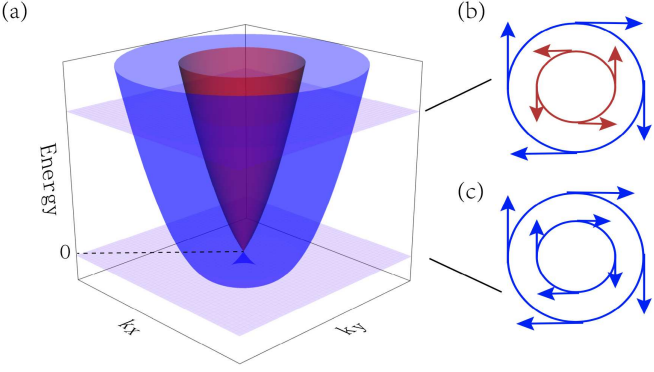}
\end{center}
\caption{(a) The conventional Rashba band structure. (b) and (c) The spin
textures of two Fermi circles when Fermi energy $E_{F}>0$ and $E_{F}<0$,
respectively. }%
\label{f1}%
\end{figure}

We start with the conventional Rashba bands described by $E_{\pm
}(k)=\varepsilon\mathbf{k}^{2}\pm\alpha_{R}|\mathbf{k}|$ with $\varepsilon$
and $\alpha_{R}$ the constant and Rashba SOC parameters, respectively. The two
outer and inner bands and the Fermi contour involving two Fermi circles for a
specific Fermi energy $E_{F}$ are shown in Fig. \ref{f1}. The spin textures of
the two Fermi circles are denoted by the red and blue arrows, as shown in Fig.
\ref{f1} (b) and (c). In the case of Fermi energy $E_{F}>0$, the two Fermi
circles have spin textures with opposite chiralities and group velocities
$\mathbf{v}_{F}^{\pm}=\frac{1}{\hslash}\nabla_{\mathbf{k}}E_{\pm
}(k)|_{\mathbf{k}=\mathbf{k}_{F}^{\pm}}$ with the same directions, as shown in
Fig. \ref{f1} (b). For $E_{F}<0$, the two Fermi circles are from the same
outer band $E_{-}(k)$, and they have the spin textures with same chiralities
but group velocities $\mathbf{v}_{F}^{-}$ with opposite directions, as shown
in Fig. \ref{f1} (c). When the spin current is injected, the opposite chiral
spin textures in the former case and the opposite directions of the velocities
in the latter case can suppress the converted charge current, because the two
Fermi circles give a partial compensation of the contributions to the
converted charge current\cite{PRL-Sn-2016}. This is the key reason to limit
the spin-to-charge conversion efficiency $\lambda_{IEE}$ in Rashba system. To
break the bottleneck, a natural strategy is to force both Fermi circles to
have positive contributions to spin-to-charge conversion. Namely, both Fermi
circles have spin textures with the same chiralities and the group velocities
with the same directions. Such unconventional Rashba band structure was
experimentally observed in some surface alloy systems such as Bi/Cu(111), and
was argued to originate from the hybridizations between different bands and
orbitals\cite{PRB-unconventional,PRB-BiTeI,kptheory-2019,Hy-twobands-2012}.

Inspired by these experimental observations, we consider two sets of Rashba
bands with couplings, under the basis [$c_{1,k\uparrow},c_{1,k\downarrow
},c_{2,k\uparrow},c_{2,k\downarrow}$]$^{T}$, the $4\times4$ Hamiltonian can be
expressed as
\begin{equation}
H(k)=H_{0}(k)+H_{c}(k),\label{H_tot}%
\end{equation}
where $c_{i,k\sigma}$ is the electron annihilation operator with $i$, $k$ and
$\sigma$ labeling the orbital, momentum and spin indexes, respectively.
$H_{0}$ describes two sets of independent Rashba bands and has the form
\begin{equation}
H_{0}(k)=\left(
\begin{array}
[c]{cccc}%
\varepsilon k^{2}+\varepsilon_{0} & i\alpha_{R}k_{-} & 0 & 0\\
-i\alpha_{R}k_{+} & \varepsilon k^{2}+\varepsilon_{0} & 0 & 0\\
0 & 0 & \varepsilon^{\prime}k^{2}+\varepsilon_{0}^{\prime} & i\alpha_{R}%
k_{-}\\
0 & 0 & -i\alpha_{R}k_{+} & \varepsilon^{\prime}k^{2}+\varepsilon_{0}^{\prime}%
\end{array}
\right)  .\label{H0}%
\end{equation}
Here, ($\varepsilon,\varepsilon_{0}$) and ($\varepsilon^{\prime}%
,\varepsilon_{0}^{\prime}$) are the parameters for the two sets of Rashba
bands, respectively, $k_{\pm}=k_{x}\pm ik_{y}$. $H_{c}(k)$ describes the
spin-flip inter-band couplings and can be parameterized as
\begin{equation}
H_{c}(k)=\left(
\begin{array}
[c]{cccc}%
0 & 0 & 0 & V_{14}(k)\\
0 & 0 & V_{23}(k) & 0\\
0 & V_{23}^{\ast}(k) & 0 & 0\\
V_{14}^{\ast}(k) & 0 & 0 & 0
\end{array}
\right)  .\label{HI}%
\end{equation}
Here we keep the terms of $H_{c}(k)$ to first order because higher-order terms
do not change the in-plane spin textures but only induces warping effect (see
Sec. III in Supplementary Materials (SMs) for details). The explicit form of
$H_{c}$ is restricted by the symmetry of the physical system we concern. We
consider a C$_{3v}$ point group which is exactly possessed by Bi/Cu(111),
Pb/Si(111), PbBiI and our predicted monolayer OsBi$_{2}$. Combined with
time-reversal (TR) symmetry, $H_{c}(k)$ should satisfy%

\begin{equation}
\hat{T}^{-1}H_{c}(k)\hat{T}=H_{c}(-k) \label{tr}%
\end{equation}
and
\begin{equation}
D(\hat{g})H_{c}(k)D^{-1}(\hat{g})=H_{c}(\hat{g}k), \label{pointgroup}%
\end{equation}
where $\hat{g}$ is the symmetry operation of point group, $D(\hat{g})$ is the
matrix representation of $\hat{g}$, and $\hat{T}$ is TR operator. For C$_{3v}$
point group, there are two group generators which are rotating $\frac{2\pi}%
{3}$ around $z$-axis and reflecting by the three vertical mirrors. Applying
these restrictions on $H_{c}(k)$, we obtain $V_{14}=V_{23}^{\ast}=\beta k_{-}%
$, where $\beta$ is a real parameter (See Sec. V in SMs for details).
\begin{figure}[ptb]
\begin{center}
\includegraphics[width=1.0\columnwidth]{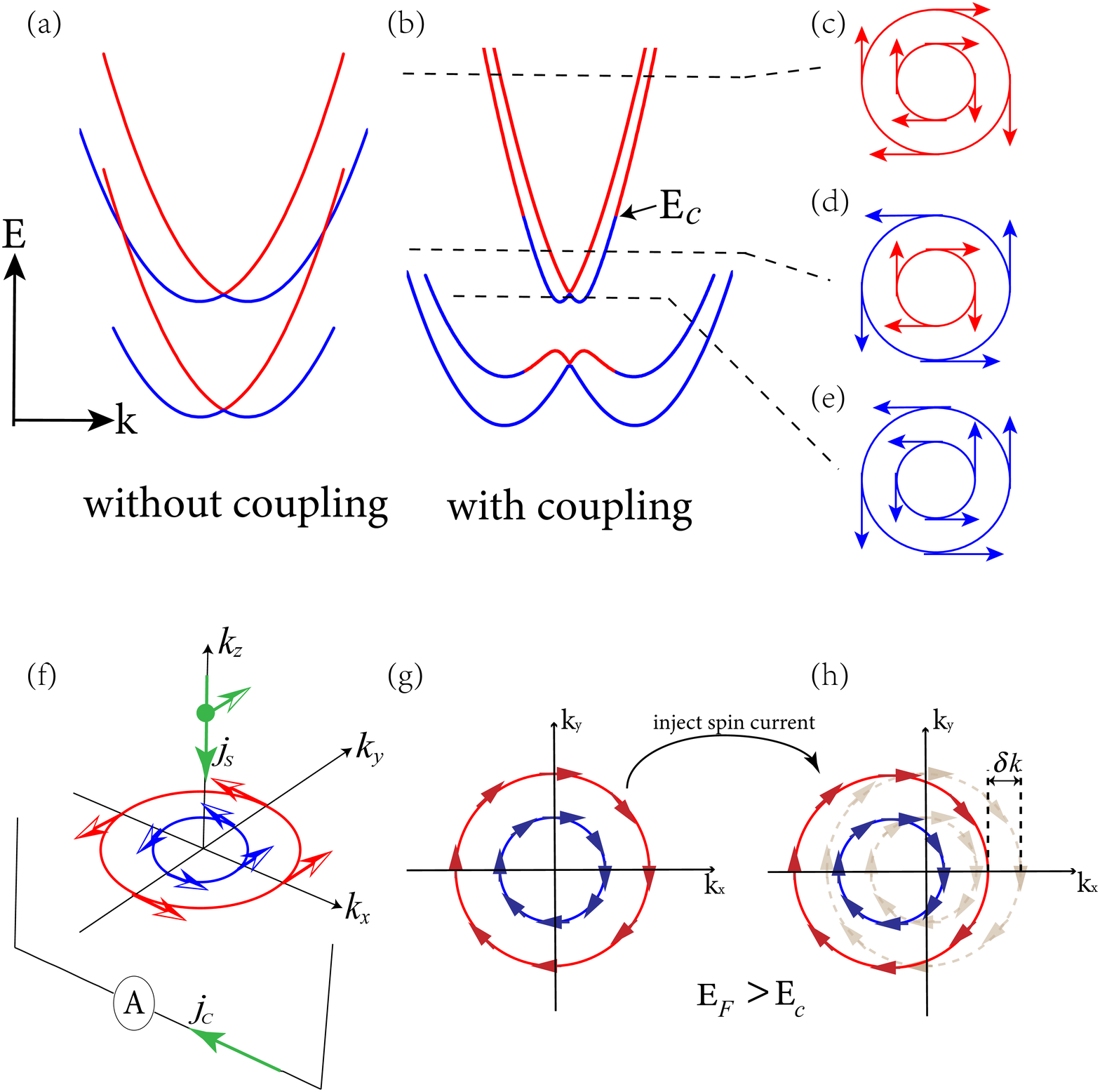}
\end{center}
\caption{(a) and (b) The two sets of Rashba bands without and with couplings,
respectively. (c), (d) and (e) The spin textures for the two Fermi circles
from the two higher-energy bands, when the Fermi energy $E_{F}$ lies in Region
III, II and I, respectively. (f) The schematic diagram for the spin-to-charge
conversion through the unconventional Rashba band structure shown in (c),
i.e., IEE. The $\hat{z}$-directional spin current $j_{s}$ with $\hat{y}%
$-directional spin polarization can generates $\hat{x}$-directional charge
current $j_{c}$. (g) and (h) The configurations of the two Fermi circles from
two higher-energy bands shown in (c) before and after injection of spin
current, respectively. }%
\label{f2}%
\end{figure}

The band structures from $H(k)$ of Eq. (\ref{H_tot}) depend on the values of
($\varepsilon^{\prime},\varepsilon_{0}^{\prime}$) relative to ($\varepsilon
,\varepsilon_{0}$). Here, we focus on the case with ($\varepsilon^{\prime
},\varepsilon_{0}^{\prime}$)=($\varepsilon,0$), which is also the situation
owned by monolayer OsBi$_{2}$. The discussions of other cases are shown in
Sec. I in SMs. Assume all the constant parameters are positive, $H(k)$ can be
solved analytically. The band structures are shown in Fig. \ref{f2}. Consider
a couple of inner and outer Fermi circles from the two higher-energy bands as
shown in Fig. \ref{f2}(b), the spin textures can be evaluated by
$\mathbf{S}_{\mathbf{k}}^{\pm}=\langle\Psi_{\pm}(\mathbf{k}%
)|\mathbf{\bm{\Omega}}|\Psi_{\pm}(\mathbf{k})\rangle$. $\mathbf{\bm{\Omega}=}%
\tau_{0}\otimes\mathbf{\bm{\sigma}}$, with $\tau_{0}$ \ and
$\mathbf{\bm{\sigma}}$ spanning the orbital and spin space. Then,%
\begin{align}
\mathbf{S}_{\mathbf{k}}^{+} &  =\frac{2\alpha_{R}k-\varepsilon_{0}}%
{\sqrt{(2\alpha_{R}k-\varepsilon_{0})^{2}+\beta^{2}k^{2}}}(\sin\theta
\mathbf{\hat{x}}-\cos\theta\mathbf{\hat{y}}),\label{sp}\\
\mathbf{S}_{\mathbf{k}}^{-} &  =\frac{2\alpha_{R}k+\varepsilon_{0}}%
{\sqrt{(2\alpha_{R}k+\varepsilon_{0})^{2}+\beta^{2}k^{2}}}(\sin\theta
\mathbf{\hat{x}}-\cos\theta\mathbf{\hat{y}}),\label{sm}%
\end{align}
according to which, three regions can be divided by different fillings. Region
I is from $E_{min}$ to zero, where the two Fermi circles is from the single
outer band and have the same chiral spin textures, as shown in Fig.
\ref{f2}(e). Region II is from zero to $E_{c}$ with $E_{c}=E^{outer}(k_{c})$
and $k_{c}=\frac{\varepsilon_{0}}{2\alpha_{R}}$. In region II, the two Fermi
circles possess opposite chiral spin textures, as shown in in Fig.
\ref{f2}(d). Region III is from $E_{c}$ to infinity, where the two Fermi
circles share the same chiral spin textures, as shown in in Fig. \ref{f2}(c).
More remarkably, two Fermi circles in Region III have the same directions of
velocities. This means the band structures in the Region III belong to the
unconventional Rashba-type. The situations are similar for the two
lower-energy bands in Fig. \ref{f2} (b). \begin{figure}[ptb]
\begin{center}
\includegraphics[width=1.0\columnwidth]{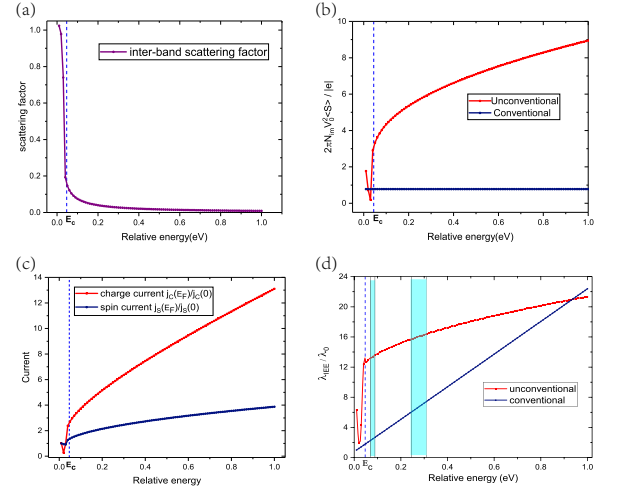}
\end{center}
\caption{(a) The inter-band scattering factor as function as Fermi energy,
which is proportional to the inter-band transition rate $|\langle\Psi
_{+}(k)|\Psi_{-}(k)\rangle|^{2}$. (b) Total spin polarization of conventional
and unconventional Rashba systems, where we adopt the same parameter of Rashba
coupling $\alpha_{R}$. (c) The relative spin current $j_{s}(E_{F})/j_{s}(0)$
and the relative charge current $j_{c}(E_{F})/j_{c}(0)$ as the function as the
Fermi energy for the unconventional Rashba systems . (d) The spin-to-charge
conversion efficiency $\lambda_{IEE}$ of the conventional and unconventional
Rashba systems as the function as the Fermi energy. Here, $\lambda_{IEE}$ is
in unit of $\lambda_{0}$, which is the IEE length of conventional Rashba
system at $E_{F}=0$. The shadowed regions correspond to the shadowed regions
shown in Fig. 4 (d). In both (c) and (d), we adopt the same fitting parameters
from monolayer OsBi$_{2}$.}%
\label{f3}%
\end{figure}

Now, we consider the spin injection process as shown in Fig. \ref{f2} (f). In
Region III, the two Fermi circles should move $\delta\mathbf{k}$ along the
same direction in momentum space, as shown in Fig. \ref{f2} (g) and (h). Such
Fermi circles shift means the in-plane charge current is generated. This is
the physical picture of the spin-to-charge conversion, \textit{i.e.}, the IEE.
In the semiclassical Boltzmann transport theory \cite{semiclassical}, the
shift of the Fermi circles is equivalent to application of a homogeneous
electrostatic field $\mathbf{E}$, which generates a directional current and
makes the distribution function $f_{k}$ to deviate from the equilibrium
distribution function $f_{k}^{0}$. In the zero-temperature limit, we have
$f_{k}=f_{k}^{0}-|e|\mathbf{\bm{\Lambda_{k}}}\cdot\mathbf{E}\ \delta
(E_{k}-E_{F})$, and the spin polarization $\left\langle \mathbf{S}%
\right\rangle $ can be expressed as $\left\langle \mathbf{S}\right\rangle
=\sum_{k}\mathbf{S}_{k}(f_{k}-f_{k}^{0})$. Here, $e$ is the elementary charge,
and $\mathbf{\bm{\Lambda_{k}}}$ is the mean free path. Under the
relaxation-time approximation, $\mathbf{\bm{\Lambda_{k}}}=\tau_{k}%
\mathbf{v}_{k}$ with $\tau_{k}$ and $\mathbf{v}_{k}$ the momentum relaxation
time and the group velocity, respectively. We consider two Fermi circles from
two higher-energy bands, as shown in Fig. \ref{f2}. (b). Note that the results
for two Fermi circles from two lower-energy bands are similar. With the help
of the spin textures in Eqs. (\ref{sp}) and (\ref{sm})(See Sec. II in SMs for
details), we have
\begin{equation}
\left\langle \mathbf{S}\right\rangle =\frac{|e|A}{2\pi\hbar}\sum\limits_{\eta
}I^{\eta}(k_{F}^{\eta})\tau_{F}^{\eta}k_{F}^{\eta}(\mathbf{\hat{v}}_{F}^{\eta
}\cdot\mathbf{\hat{k}}_{F}^{\eta})(\mathbf{\hat{z}}\times\mathbf{E}%
).\label{simplify}%
\end{equation}
Here, $\eta=\pm$ labels the inner and outer Fermi circles, respectively.
$I^{\eta}(k)=(2\alpha_{R}k+\eta\varepsilon_{0})/\sqrt{(2\alpha_{R}%
k+\eta\varepsilon_{0})^{2}+\beta^{2}k^{2}}$ is the factor for spin
polarization. $A,\mathbf{\hat{v}}_{F}$ and $\mathbf{\hat{k}}_{F}$ is the area
of the unit cell, the unit vector of group velocity and Fermi momentum,
respectively. Since the two Fermi circles share identical directions of spin
polarization and group velocity in Region III, they both give positive
contributions to total spin polarization. Additionally, as shown in Fig.
\ref{f3}(a), the rate of inter-band scattering is greatly reduced with the
increase of $E_{F}$, because the spin-flip backscattering is forbidden.
Accordingly, the momentum relaxation time is also increased. These two aspects
strongly enhance the spin polarization $\left\langle \mathbf{S}\right\rangle
$, as shown in Fig. \ref{f3}(b). When $E_{F}\gg E_{c}$, the factor of spin
polarization and radii of two Fermi circles tend to be equal, the spin
polarization can be approximately expressed as
\begin{equation}
\left\langle \mathbf{S}\right\rangle _{E_{F}\gg E_{c}}=\frac{\alpha_{R}%
|e|}{2\pi N_{im}V_{0}^{2}}(\sqrt{1+\frac{4\varepsilon\Delta E_{F}}{\alpha
_{R}^{2}+\beta^{2}}}-1)(\mathbf{\hat{z}}\times\mathbf{E}),\label{aapoS}%
\end{equation}
with $\Delta E_{F}=E_{F}-\frac{1}{2}\varepsilon_{0}$. Here, $N_{im}$ and
$V_{0}$ denote the number of $\delta$-scattering centers and the s-wave
scattering potential, respectively. In conventional Rashba system, the spin
polarization $\left\langle \mathbf{S}\right\rangle _{Rashba}=\alpha
_{R}|e|/(2\pi N_{im}V_{0}^{2})$ (See Sec. II in SMs for details), which is a
constant when $E_{F}>0$. Thus, the spin polarization in unconventional Rashba
system rapidly surpasses the conventional one, as shown in Fig. \ref{f3}(b).

The spin current density $\mathbf{j}_{s}$ shown in Fig. \ref{f3} (c) can be
related with the spin polarization $\left\langle \mathbf{S}\right\rangle $
shown in Eq. (\ref{simplify}) by $\mathbf{j}_{s}=\frac{e\left\langle
S\right\rangle }{\tau_{F}}\mathbf{\hat{z}}$. The generated charge current
density $\mathbf{j}_{c}$ shown in Fig. \ref{f3} (c) can be obtained by
$\mathbf{j}_{c}=e\sum_{k}\mathbf{v}_{k}(f_{k}-f_{k}^{0})$. The spin-to-charge
conversion efficiency $\lambda_{IEE}$ can be further expressed as%

\begin{equation}
\lambda_{IEE}=j_{c}/j_{s}=\frac{\sum_{\eta}k_{F}^{\eta}\tau_{F}^{\eta}%
v_{F}^{\eta}}{\sum_{\eta}I^{\eta}(k_{F}^{\eta})k_{F}^{\eta}(\mathbf{\hat{v}%
}_{F}^{\eta}\cdot\mathbf{\hat{k}}_{F}^{\eta})}. \label{ratio}%
\end{equation}
For the conventional Rashba system, $\lambda_{IEE}^{con}\sim\alpha_{R}\tau
_{F}/\hslash+4\varepsilon\tau_{F}E_{F}/(\alpha_{R}\hslash)$ with a nearly
constant $\tau_{F}$. Then, $\lambda_{IEE}^{con}$ tends to be the constant
$\lambda_{0}$, when $E_{F}\sim0$, and is linearly increased with the shift of
$E_{F}$, indicated by the blue line in Fig. \ref{f3}(d). For the
unconventional Rashba system, however, with the increase of $E_{F}$, the
inter-band scattering rate $\propto$ $|\langle\Psi_{+}(k)|\Psi_{-}%
(k)\rangle|^{2}$ rapidly decreases, as shown in Fig. \ref{f3}(a), which gives
a large momentum-relaxation time $\tau_{F}$. Thus, the larger charge current
$j_{c}$ is generated. Meanwhile, stronger spin polarization generates larger
spin current $j_{s}$. These two effects compete with each other. The numerical
results of the relative $j_{c}$ and $j_{s}$ for the unconventional Rashba
system are shown in Fig. \ref{f3}(c), from which, one can clearly find the
competitive relation between $j_{c}$ and $j_{s}$. Compared with conventional
Rashba system, unconventional Rashba system has lower rate of increase but
much larger initial value of $\lambda_{IEE}$ length, as indicated by the red
curve in Fig. \ref{f3}(d). In the energy window we most concern, such as the
shadowed regions in Fig. \ref{f3}(d), which correspond to the situations in
monolayer OsBi$_{2}$ with $\lambda_{IEE}^{ucon}/\lambda_{IEE}^{con}\sim10$.
The unconventional Rashba system has remarkable advantages than conventional
Rashba systems. \begin{figure}[ptb]
\begin{center}
\includegraphics[width=1.0\columnwidth]{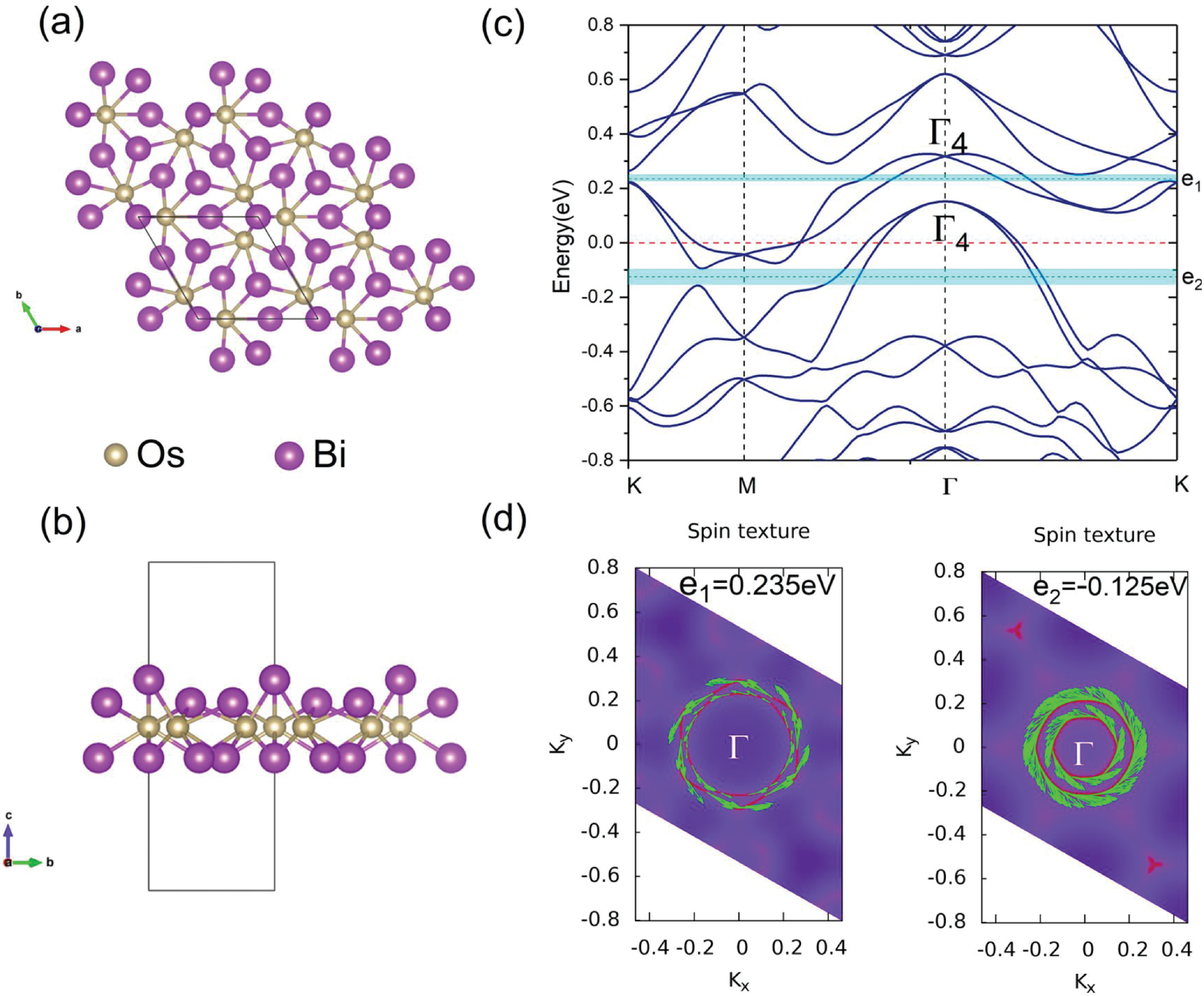}
\end{center}
\caption{(a) and (b) The top and side view of the Crystal structure of
monolayer OsBi$_{2}$, respectively. (c) The band structures along the
high-symmetry lines. (d) and (e) The spin textures of different Fermi circles
around $\Gamma$ point for two different Fermi energy 0.235eV and -0.125eV,
respectively. The green arrows indicate the direction and intensity of spin
textures.}%
\label{f4}%
\end{figure}

The search for a simple physical material with the unconventional Rashba band
structures is of fundamental importance for achieving the giant IEE. Our
previous studies indicate the trigonal layered PtBi$_{2}$-type materials tend
to form the buckled structure for the top layer of Bi \cite{bulk-PtBi2}. Such
kinds of distortion naturally breaks the inversion symmetry. Following the
similar strategy, we perform the first-principles calculations to search for
the same structures with different elements (See Sec. IV in SMs for details).
We find that the monolayer OsBi$_{2}$ meets all the requirements. The relaxed
structure is shown in Fig. \ref{f4}(a) and (b). The crystal constant is
$a=b=6.78${\AA }. The calculated band structure is shown in Fig. \ref{f4}(c).
The point group is C$_{3v}$ as our aforementioned discussions. With SOC, all
the bands at $\Gamma$ point can be classified into three Irreducible
representations (IRs), $\Gamma_{4}$, $\Gamma_{5}$ and $\Gamma_{6}$
\cite{group}. Deviate from $\Gamma$ point, the two sets of the twofold
degenerate $\Gamma_{4}$ bands splits into Rashba-type and couple together to
give the unconventional spin textures, as conformed by Fig. \ref{f4} (d) and
(e). The details of the construction of the effective k$\cdot$p model near
$\Gamma$ point are present in Sec. V in SMs.

Note that the aforementioned discussions focus on either the two higher-energy
bands or the two lower-energy bands, as shown in Fig. \ref{f2}(b). In
continuous model described by Eq.(\ref{H_tot}), all the four bands should be
considered. However, in monolayer OsBi$_{2}$, the lattice symmetries force
only two bands to be occupied and the other two bands to be gapped when the
Fermi energy lies in the shadowed regions, as shown in Fig. \ref{f4} (c). This
is a crucial point to guarantee the aforementioned results can be applied in
monolayer OsBi$_{2}$. With the fitting parameters from monolayer OsBi$_{2}$,
the $\lambda_{IEE}$ of OsBi$_{2}$ in the shadowed regions indeed realize the
giant IEE with $\lambda_{IEE}^{ucon}/\lambda_{IEE}^{con}\sim10$, as indicated
in Fig. \ref{f3} (d). The Fermi level of the pristine monolayer OsBi$_{2}$
does not lie in the shadowed region in Fig. \ref{f4}(b). To tune it, many
practical methods can be adopted, such as chemical doping, electrostatic
gating and ionic liquid gating
etc\cite{Gate-1,Gate-2,Gate-3,Gate-4,Gate-5,Gate-6,Gate-7}.

In conclusion, we propose that the unconventional Rashba system can give rise
to a new mechanism to realize the spin-to-charge conversion. With the help of
semiclassical transport theoretical analysis, we find that the spin-to-charge
conversion efficiency is much higher than that in the conventional Rashba
system. More meaningfully, our first-principles calculations prove that
monolayer OsBi$_{2}$ belongs to such unconventional Rashba system. It makes
monolayer OsBi$_{2}$ to be a promising material to host the giant IEE. Our
studies provide a new mechanism to promote the charge-to-spin conversion
efficiency in modern spintronics.

\begin{acknowledgments}
This work was financially supported by the National Key R\&D Program of China
No. 2017YFA0303201, National Natural Science Foundation of China under Grants
(No. 12022413, No. 11674331, No.11625415), the \textquotedblleft Strategic
Priority Research Program (B)\textquotedblright\ of the Chinese Academy of
Sciences, Grant No. XDB33030100, the `100 Talents Project' of the Chinese
Academy of Sciences, the Collaborative Innovation Program of Hefei Science
Center, CAS (Grants No. 2020HSC-CIP002), the CASHIPS Director's Fund
(BJPY2019B03), the Science Challenge Project under Grant No. TZ2016001. A
portion of this work was supported by the High Magnetic Field Laboratory of
Anhui Province, China.
\end{acknowledgments}


\begin{thebibliography}{99}                                                                                               %


\bibitem {PRM-spin-hall}J. Sinova, Sergio O. Valenzuela, J. Wunderlich, C. H.
Back, and T. Jungwirth, Rev. Mod. Phys. \textbf{87}, 1213 (2015).

\bibitem {spintronics-2004}I. \v{Z}uti\'{c}, J. Fabian, and S. Das Sarma, Rev.
Mod. Phys. \textbf{76}, 323 (2004).

\bibitem {Edelstein-1989}V. M. Edelstein, Solid State Commun. \textbf{73}, 233 (1990).

\bibitem {PRL-2007-metal}C. R. Ast, J. Henk, A. Ernst, L. Moreschini, M. C.
Falub, D. Pacil\'{e} P. Bruno, K. Kern, and M. Grioni, Phys. Rev. Lett.
\textbf{98}, 186807 (2007).

\bibitem {NC-2013-metal}J. C. Rojas-Sanchez, L. Vila, G. Desfonds, S.
Gambarelli, J. P. Attane, J. M. D. Teresa, C. Magen, and A. Fert, Nat. Commun.
\textbf{4}, 2944 (2013).

\bibitem {APL-2015-metal1}S. Sangiao, J. M. D. Teresa, L. Morellon, I. Lucas,
M. C. Martinez-Valarte, and M. Viret, Appl. Phys. Lett. \textbf{106}, 172403 (2015)

\bibitem {APL-2015-metal2}A. Nomura, T. Tashiro, H. Nakayama, and K. Ando,
Appl. Phys. Lett. \textbf{106}, 212403 (2015).

\bibitem {PRB-2016-metal3}M. B. Jungfleisch, W. Zhang, J. Sklenar, W. Jiang,
J. E. Pearson, J. B. Ketterson, and A. Hoffmann, Phys. Rev. B \textbf{93},
224419 (2016).

\bibitem {PRB-2016-metal4}M. Isasa, M. C. Mart\'{\i}nez-Valarte, E. Villamor,
C. Mag\'{e}n, L. Morell\'{o}n, J. M. De Teresa, M. R. Ibarra, G. Vignale, E.
V. Chulkov, E. E. Krasovskii, L. E. Hueso, and F. Casanova, Phys. Rev. B
\textbf{93}, 014420 (2016).

\bibitem {PRL-2015-metal}H. J. Zhang, S. Yamamoto, B. Gu, H. Li, M. Maekawa,
Y. Fukaya, and A. Kawasuso, Phys. Rev. Lett. \textbf{114}, 166602 (2015)

\bibitem {PRL-metal-2016}H. Nakayama, Y. Kanno, H. An, T. Tashiro, S. Haku, A.
Nomura, and K. Ando, Phys. Rev. Lett \textbf{117}, 116602 (2016).

\bibitem {Edelstein-1}M. Offidani, M. Milletar, R. Raimondi, and A. Ferreira,
Phys. Rev. Lett \textbf{119}, 196801 (2017).

\bibitem {Edelstein-2}L. A. Ben\'{\i}ez, W. S. Torres, J. F. Sierra, M.
Timmermans, J. H. Garcia, S. Roche, M. V. Costache and S. O. Valenzuela, Nat.
Mater. \textbf{19}, 170-175 (2020).

\bibitem {NM-1}E. Lesne, Y. Fu, S. Oyarzun, J. C. Rojas-S\'{a}nchez, D. C.
Vaz, H. Naganuma, G. Sicoli, J.-P. Attan\'{e} M. Jamet, E. Jacquet, J.-M.
George, A. Barth\'{e}l\'{e}my, H. Jaffr\`{e}s, A. Fert, M. Bibes and L. Vila,
Nat. Mater. \textbf{15}, 1261-1266 (2016).

\bibitem {PRL-Sn-2016}J.-C. Rojas-Sanchez, S. Oyarz\'{u}n, Y. Fu, A. Marty, C.
Vergnaud, S. Gambarelli, L. Vila, M. Jamet, Y. Ohtsubo, A. Taleb-Ibrahimi, P.
Le F\`{e}vre, F. Bertran, N. Reyren, J.-M. George, and A. Fert, Phys. Rev.
Lett \textbf{116}, 096602 (2016).

\bibitem {PRL-2007-TI}L. Fu, C. L. Kane, and E. J. Mele, Phys. Rev. Lett.
\textbf{98}, 106803 (2007).

\bibitem {PRB-2007-TI}L. Fu and C. L. Kane Phys. Rev. B \textbf{76}, 045302 (2007).

\bibitem {RMP-2010-TI}M. Z. Hasan and C. L. Kane, Rev. Mod. Phys. \textbf{82},
3045 (2010).

\bibitem {RMP-2011-TI}X.-L. Qi and S.-C. Zhang, Rev. Mod. Phys. \textbf{83},
1057 (2011).

\bibitem {PRB-TI-2017}M. Rodriguez-Vega, G. Schwiete, J. Sinova, and E. Rossi,
Phys. Rev. B \textbf{96}, 235419 (2017).

\bibitem {IEE-TI-2018}R. Dey, N. Prasad, L. F. Register, and S. K. Banerjee,
Phys. Rev. B \textbf{97}, 174406 (2018).

\bibitem {Rashba}Y. A. Bychkov and E. I. Rashba, J. Phys. C \textbf{17}, 6039 (1984).

\bibitem {NM-2011-SOC}K. Ishizaka, M. S. Bahramy, H. Murakawa, M. Sakano, T.
Shimojima, T. Sonobe, K. Koizumi, S. Shin, H. Miyahara, A. Kimura, K.
Miyamoto, T. Okuda, H. Namatame, M. Taniguchi, R. Arita, N. Nagaosa, K.
Kobayashi, Y. Murakami, R. Kumai, Y. Kaneko, Y. Onose and Y. Tokura, Nat.
Mater. \textbf{10}, 521-526 (2011).

\bibitem {Giant-Rashba1}J. Park, S. W. Jung, M.-C. Jung, H. Yamane, N. Kosugi,
and H. W. Yeom, Phys. Rev. Lett \textbf{110}, 036801 (2013).

\bibitem {Giant-Rashba2}A. Varykhalov, D. Marchenko, M. R. Scholz, E. D. L.
Rienks, T. K. Kim, G. Bihlmayer, J. S\'{a}nchez-Barriga, and O. Rader, Phys.
Rev. Lett \textbf{108}, 066804 (2012).

\bibitem {Giant-Rashba3}D. Niesner, M. Wilhelm, I. Levchuk, A. Osvet, S.
Shrestha, M. Batentschuk, C. Brabec, and T. Fauster, Phys. Rev. Lett
\textbf{117}, 126401 (2016).

\bibitem {Giant-Rashba4}A. Crepaldi, L. Moreschini, G. Autes, C.
Tournier-Colletta, S. Moser, N. Virk, H. Berger, Ph. Bugnon, Y. J. Chang, K.
Kern, A. Bostwick, E. Rotenberg, O. V. Yazyev, and M. Grioni, Phys. Rev. Lett
\textbf{109}, 096803 (2012).

\bibitem {Micro-theory}Ka Shen, G. Vignale, and R. Raimondi, Phys. Rev. Lett.
\textbf{112}, 096601 (2014).

\bibitem {PRB-theoretical-aspect}A. Johansson, J. Henk, and I. Mertig, Phys.
Rev. B \textbf{93}, 195440 (2016).

\bibitem {Quan-confin-1}M. S. Bahramy, P. D .C. King, A. de la Torre, J.
Chang, M. Shi, L. Patthey, G. Balakrishnan, Ph. Hofmann, R. Arita, N. Nagaosa
and F. Baumberger. Nat. Commun. \textbf{3}, 1159 (2012).

\bibitem {PRB-unconventional}H. Mirhosseini, J. Henk, A. Ernst, S. Ostanin,
C.-T. Chiang, P. Yu, A. Winkelmann, and J. Kirschner, Phys. Rev. B
\textbf{79}, 245428 (2009).

\bibitem {PRB-BiTeI}C. Mera Acosta, O. Babilonia, L. Abdalla, and A. Fazzio,
Phys. Rev. B \textbf{94}, 041302 (2016).

\bibitem {kptheory-2019}I. A. Nechaev and E. E. Krasovskii, Phys. Rev. B
\textbf{100}, 115432 (2019).

\bibitem {Hy-twobands-2012}H. Bentmann, S. Abdelouahed, M. Mulazzi, J. Henk,
and F. Reinert, Phys. Rev. Lett \textbf{108}, 196801 (2012).

\bibitem {semiclassical}I. Mertig, Rep. Prog. Phys. \textbf{62}, 237 (1999).

\bibitem {bulk-PtBi2}W. Gao, X. Zhu, F. Zheng, M. Wu, J. Zhang, C. Xi, P.
Zhang, Y. Zhang, N. Hao, W. Ning M. Tian, Nat. Commun. \textbf{9}, 3249 (2018).

\bibitem {group}G. F. Koster, \textit{Properties of the Thirty-Two Point
Groups} (MIT Press, Cambridge, MA, 1963), Vol. 24.

\bibitem {Gate-1}L. J. Li, E. C. T. O'Farrell K. P. Loh G. Eda, B.
\"{O}zyilmaz and A. H. Castro Neto, Nature \textbf{529}, 185--189 (2016).

\bibitem {Gate-2}Y. Saito, T. Nojima, Y. Iwasa, Supercond. Sci. Technol.
\textbf{29}, 093001 (2016).

\bibitem {Gate-3}K. Ueno, S. Nakamura, H. Shimotani, A. Ohtomo, N. Kimura, T.
Nojima, H. Aoki, Y. Iwasa, M. Kawasaki, Nat. Mater. \textbf{7}, 855--858 (2008).

\bibitem {Gate-4}J. M. Lu, O. Zheliuk, I. Leermakers, N. F. Q. Yuan, U.
Zeitler, K. T. Law, J. T. Ye, Science \textbf{350}, 1353--1357 (2015).

\bibitem {Gate-5}D. Costanzo, S. Jo, H. Berger, Nat. Nanotechnol. \textbf{11},
339--344 (2016).

\bibitem {Gate-6}E. Sajadi, T. Palomaki, Z. Fei, W. Zhao, P. Bement, C. Olsen,
S. Luescher, X. D. Xu, J. A. Folk, and D. H. Cobden, Science \textbf{362}, 922 (2018).

\bibitem {Gate-7}V. Fatemi, S. F. Wu, Y. Cao, L. Bretheau, Q. D. Gibson, K.
Watanabe, T. Taniguchi, T. J. Cava, and P. Jarillo-Herrero, Science
\textbf{362}, 926 (2018).
\end{thebibliography}
\end{document}